\documentclass{emulateapj}
\usepackage{natbib}

\shortauthors{T. Hosokawa et al.}
\shorttitle{Second-Generation Primordial Stars}

\newcommand{\hii}{H{\sc ii} }
\newcommand{\msun}{M_{\odot}}
\newcommand{\msunyr}{M_\odot~{\rm yr}^{-1}}
\newcommand{\mdot}{\dot{M}_*}

\begin{document}

\title{Protostellar Feedback and Final Mass \\
of the Second-Generation Primordial Stars}

\author{Takashi Hosokawa\altaffilmark{1,2},
        Naoki Yoshida\altaffilmark{1,3},
        Kazuyuki Omukai\altaffilmark{4},
        Harold W. Yorke\altaffilmark{2}}

\altaffiltext{1}{Department of Physics, University of Tokyo, 
Tokyo 113-0033, Japan;
takashi.hosokawa@phys.s.u-tokyo.ac.jp, hosokwtk@gmail.com}
\altaffiltext{2}{Jet Propulsion Laboratory, California Institute
of Technology, Pasadena CA 91109, USA}
\altaffiltext{3}{Kavli Institute for the Physics and Mathematics
of the Universe, University of Tokyo, Kashiwa, Chiba
277-8583, Japan}
\altaffiltext{4}{Department of Physics, Kyoto University, 
Kyoto 606-8502, Japan}


\begin{abstract}
The first stars in the universe ionized the ambient 
primordial gas through various feedback processes. 
``Second-generation'' primordial stars potentially form from this
disturbed gas after its recombination.
In this {\it Letter}, we study the late formation stage of
such second-generation stars, where
a large amount of gas accretes onto the protostar and  
the final stellar mass is determined when the accretion 
terminates. We directly compute the complex interplay between the
accretion flow and stellar ultraviolet (UV) radiation, performing
radiation-hydrodynamic simulations coupled with 
stellar evolution calculations.
Because of more efficient H$_2$ and HD cooling in the pre-stellar stage, 
the accretion rates onto the star are ten times lower than 
in the case of the formation of the first stars.
The lower accretion rates and envelope density result
in the occurrence of an expanding bipolar \hii region at a lower
protostellar mass $M_* \simeq 10M_{\odot}$, which blows out the 
circumstellar material, thereby quenching the mass supply from 
the envelope to the accretion disk.
At the same time the disk loses mass due to photoevaporation
by the growing star.
In our fiducial case the stellar UV feedback terminates
mass accretion onto the star at $M_* \simeq 17~M_\odot$. 
Although the derived masses of the second-generation 
primordial stars are systematically lower than those of the 
first generation, the difference is within a factor of only a few.
Our results suggest a new scenario, whereby
the majority of the primordial stars are born as massive stars 
with tens of solar masses, regardless of their generations.
\end{abstract}

\keywords{cosmology: theory -- early universe -- stars: formation -- 
stars: evolution -- accretion -- H{\sc ii} regions}

\section{Introduction}
\label{sec:intro}

\setcounter{footnote}{0}

Standard cosmology predicts that the cradles of 
the very first stars in the universe are 
massive ($100 - 1000~M_\odot$) gas clumps 
formed in minihalos hundreds million years after the Big Bang 
\citep[e.g.,][]{ABN02,BCL02,Y03}.
The massive primordial stars ionize their natal clouds 
by ultraviolet (UV) radiation.
If these stars die without triggering supernova explosions, 
their relic \hii regions, where the ionized gas recombined 
subsequently, become a promising site for the formation of
second-generation stars \citep[e.g.,][]{OH03}.
The pre-stellar thermal evolution of these 
second-generation stars differs from that of the very first stars.
The enhanced free electrons, catalysts for forming
H$_2$ molecules, in the relic \hii regions 
results in the higher H$_2$ abundance than in the case of 
the first stars.
The higher H$_2$ abundance enables the production of 
cold ($\lesssim 150$~K) dense gas, where 
most of the deuterium is converted to molecular form
\citep[e.g.,][]{UI00,NU02,NO05,JB06}
by the exothermic reaction 
\begin{equation}
{\rm D^{+} + H_2 \rightarrow HD + H^{+}}.
\end{equation}
With the additional radiative cooling via HD molecules, 
the temperature further falls to a few $\times 10$~K, which is
close to the CMB temperature at high redshifts. 
By contrast, in the case of first-star formation 
the temperature remains $\ga$ 200K. 
Strong shocks or cosmic-rays 
can raise the ionization degree of a primordial gas,
thereby setting similar physical conditions
\citep[e.g.,][]{SB07,Safr10,IO11,O12}.
The typical mass of the star-forming clouds is given by the Jeans
mass when the contracting gas attains its minimum temperature
\begin{equation}
M_{\rm c} \sim M_J \simeq 70~M_\odot 
\left( \frac{n}{10^5 {\rm cm}^{-3}} \right)^{-1/2}
\left( \frac{T}{50~{\rm K}} \right)^{3/2},
\end{equation}
where the characteristic density and temperature are set by 
the HD cooling \citep[e.g.,][]{Osh05,YOKH07,MB08}. 
This cloud mass-scale  
is about an order of magnitude 
less massive than that for the formation of the very first stars:
\begin{equation}
M_{\rm c} \sim M_J \simeq 2000~M_\odot 
\left( \frac{n}{10^4 {\rm cm}^{-3}} \right)^{-1/2}
\left( \frac{T}{200~{\rm K}} \right)^{3/2},
\end{equation}
where the characteristic density and temperature are set by 
the H$_2$ cooling \citep{BCL99}.

Because of the difference in the mass-scale of their natal clouds, 
the first- and second-generation primordial 
stars have been classified into different sub-populations: 
Population III.1 (Pop III.1) and III.2
(Pop III.2) stars \citep[e.g.,][]{Osh08}.
Until recently, the standard picture of primordial star formation 
postulated very massive (a few $\times 100~M_\odot$) 
Pop III.1 stars and less massive Pop III.2 stars 
(a few $\times 10~M_\odot$). 
However, their final stellar masses are still not clear.
Both in the Pop III.1 and III.2 cases, very small ($\simeq 0.01~M_\odot$) 
embryo protostars form after the gravitational collapse of the natal clouds
\citep[e.g.,][]{ON98,YOH08}. These protostars then rapidly grow in mass 
by accretion of gas from their surrounding envelopes.
The final stellar mass is thought to be set when the mass accretion
ceases.

The key physical mechanism that controls the final stellar mass
is stellar radiation feedback against the accretion flow
\citep[e.g.,][]{MT08,HOYY11,Stacy12}.
In our previous work \citep[][hereafter H11]{HOYY11}, 
we have reported the first radiation-hydrodynamic simulations 
for the Pop III.1 case that cover the long-term ($\simeq 10^5$~years) 
evolution after the birth of the embryo protostar.
We have demonstrated that the strong stellar UV feedback indeed terminates 
protostellar accretion, and shown that the resulting final 
stellar mass is $\simeq 40~M_\odot$.
This mass range of Pop III.1 stars is much lower than previously 
postulated and is rather close to that expected for Pop III.2 stars.

In this {\it Letter}, we apply the same method to a typical
Pop III.2 star-forming cloud found in a cosmological simulation 
\citep{YOH07}. 
We find that although the final masses of Pop III.2 stars are
systematically lower than those of Pop III.1 stars, the 
difference is by a factor of only a few.
This result suggests that, whereas lower-mass stars potentially 
form as satellite-like objects via gravitational fragmentation 
of the star-forming cloud \citep[e.g.,][]{Mcd08,Turk09,Stacy10,Cl11}, 
the majority of primordial stars are 
tens of solar masses regardless of their generations.
The typical stellar systems in the early universe are likely to be 
such massive stars accompanied by a small number
lower-mass satellite stars.

\section{Numerical Method}
\label{sec:method}

\begin{deluxetable}{llc}
\tablecaption{Included Deuterium chemistry \label{tab:reactions}}
\tablehead{\colhead{No.} & \colhead{Reactions} & \colhead{References}}
\startdata
R1 &  D$^+$  +  e  $\rightarrow$ D + $\gamma$  &  1  \\
R2 &  D  +  H$^+$  $\rightarrow$ D$^+$ + H     &  1  \\
R3 &  D$^+$  +  H  $\rightarrow$ D + H$^+$     &  1  \\
R4 &  D + H $\rightarrow$ HD + $\gamma$        &  2  \\
R5 &  D + H$_2$  $\rightarrow$  H + HD         &  3  \\
R6 &  D$^+$  +  H$_2$  $\rightarrow$  H$^+$ + HD  &  2  \\
R7 &  HD + H  $\rightarrow$  H$_2$ + D            &  2  \\
R8 &  HD  + H$^+$ $\rightarrow$  H$_2$ + D$^+$    &  2  \\
R9 &  HD  + $\gamma$ $\rightarrow$  H + D   &  4 \\
\enddata
\tablecomments{1: \citet{GP98}, 2: \citet{Stancil98}, 3: \citet{WS02},
4: \citet{WGH11}}
\end{deluxetable}

We employ the numerical method developed in H11 with necessary extensions.
The hydrostatic evolution of the central protostar and the evolution of 
the accreting primordial gas are consistently solved (see H11 for details). 
The dynamics of the accretion flow is calculated with a two-dimensional (2D)
axisymmetric radiation-hydrodynamics code \citep[e.g.,][]{YS02,YW96}. 
In addition to the chemical reactions implemented in H11,
we also solve the deuterium chemistry with three species
of D, HD, and D$^+$ (Table 1).
The HD-line cooling rate is calculated considering 
the first four rotational levels as in \citet{GP98}.
The central protostar is replaced with a sink cell,
whose size is $\simeq 10$~AU.
This masking of the very vicinity of the star,
where the gas density exceeds $10^{11}~{\rm cm}^{-3}$,
enables us to follow the long-term evolution.
The mass accretion rate onto the protostar is evaluated 
from the mass influx to the sink cell.
With the accretion rate provided, the protostellar evolution
is calculated by solving the stellar 
interior structure numerically \citep[e.g.,][]{OP03,HO09}.
The stellar evolution then gives the stellar luminosity
and effective temperature to the radiation hydrodynamics, 
which follows the interplay between the accretion flow and
stellar radiation.

As in H11 we calculate the dynamics of the accretion flow within
0.3~pc of the protostar.
The initial gas distribution is configured with the simulation output of
\cite{YOH07},
who study the formation of a protostellar core 
in a relic \hii region, performing three-dimensional (3D) 
cosmological simulations. 
We reduce their 3D data at the moment 
when the central density reaches $\simeq 10^{14}~{\rm cm}^{-3}$
to 2D axisymmetric data by averaging over azimuthal angles. 
The original 3D data has some non-axisymmetric structure, but
the deviation from the axial average is not large within the 
computational domain. Moreover, deviation from the axi-symmetric 
structure is smaller in the inner part of the cloud.
Our computational domain contains $\simeq 150~M_\odot$.
\citet{YOKH07,YOH07} show that in the Pop III.2 case the mass of 
the cloud that experiences the dynamical run-away collapse is 
about $40~M_\odot$. 
In our initial condition, this part of 
the cloud is contained within 0.05~pc of the protostar.
In comparison with the Pop III.1 case (H11), the density
in the accreting envelope is lower,
reflecting the lower temperature in the pre-stellar collapse 
stage. The grid resolution and boundary conditions are
the same as in the fiducial case studied in H11.

\section{Results}
\label{sec:results}

Immediately after the onset of the calculation, the stellar mass 
increases rapidly by accretion via a circumstellar disk.
The density within the disk is high enough 
($\ga 10^8~{\rm cm}^{-3}$) for the three-body reaction 
to convert most of the H atoms to H$_2$ molecules.
This molecular disk extends to $\simeq 500$~AU from the star
when the stellar mass reaches $5~M_\odot$. 
The mass accretion rate onto the protostar decreases with increasing 
stellar mass, e.g., $1.7 \times 10^{-3}~\msunyr$ at 
$M_* = 2~M_\odot$, while $6.1 \times 10^{-4}~\msunyr$ at
$M_* = 5~M_\odot$.
These rates are about an order of magnitude 
lower than those in the Pop III.1 case,
due to the lower density in the accreting envelope
resulting from the lower gas temperature during the run-away
collapse stage: for the collapse of a self-gravitating cloud, 
the accretion rate is given by 
$\mdot \sim c_s^3 / G \propto T^{1.5}$,
where $c_s$ and $T$ are the sound speed and gas temperature 
in the accreting envelope \citep{Shu77}.
In the Pop III.2 case, as described above, the gas temperature 
is lower than in the Pop III.1 case because of the efficient 
radiative cooling via extra H$_2$ and HD molecules.

\begin{figure}
  \begin{center}
\epsscale{1.0}
\plotone{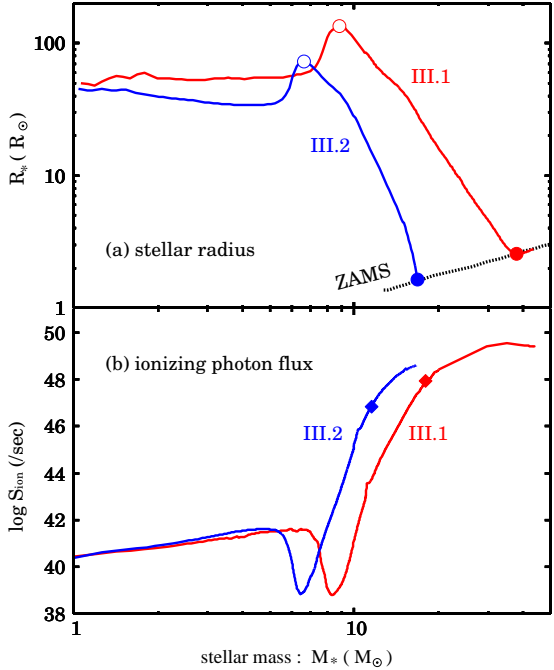}
\caption{Evolution of the protostellar radius ({\it upper panel}) and 
ionizing ($h \nu > 13.6$~eV) photon production rate ({\it lower panel}).
The blue and red lines represent the fiducial Pop III.2 and 
III.1 cases in each panel. The black dotted line in the upper panel 
shows the mass-radius relation of metalless zero-age main-sequence stars. 
The open and filled circles mark the characteristic 
epochs of the protostellar evolution, beginning of the KH contraction 
and the arrival of the protostar to the ZAMS.
The filled diamonds in the lower panel denote the epochs when the
\hii region begins to grow toward the polar directions of the disk. 
}
\label{fig:mr_suv}
  \end{center}
\end{figure}

Figure~\ref{fig:mr_suv}-(a) shows that, early on, 
the protostellar radius is $30 - 40~R_\odot$ and decreases slightly 
as the accretion rate decreases. 
The protostellar evolution can be understood by comparing 
the two timescales, namely the stellar Kelvin-Helmholtz (KH) timescale
\begin{equation}
t_{\rm KH} \equiv \frac{G M_* \mdot}{R_* L_*} ,  
\end{equation}
and the accretion timescale
\begin{equation}
t_{\rm acc} \equiv \frac{M_* }{\mdot}.  
\end{equation} 
The former is the times over which the star radiates away 
its internal heat content, whereas the latter
is the time over which the stellar mass doubles by accretion. 
Initially, the KH timescale is much longer 
than the accretion timescale, i.e., the star has no time to 
radiate its internal heat before its mass grows by accretion. 
This is the so-called adiabatic accretion stage \citep[e.g.,][]{SPS86}.
Since the stellar luminosity $L_*$ rises rapidly 
with increasing stellar mass, the KH timescale eventually falls 
below the accretion timescale at $M_* \simeq 6~M_\odot$, 
where the stellar radius marks its maximum. 
After this the protostar loses its internal heat 
efficiently by radiative diffusion and contracts quasi-hydrostatically
\citep[e.g.,][]{OP03}. 
During this KH contraction stage the stellar effective temperature
$T_{\rm eff}$ varies as $\propto R_*^{-2}$ and 
H ionizing-photon ($h \nu \geq 13.6$~eV) luminosity therefore 
increases significantly (Fig.~\ref{fig:mr_suv}-b).

\begin{figure}
  \begin{center}
\epsscale{1.1}
\plotone{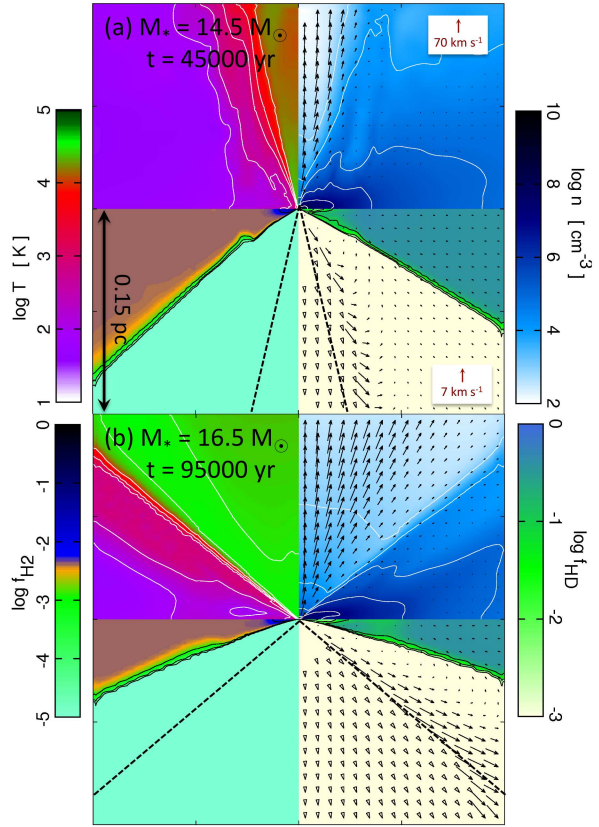}
\caption{The structure of the accreting envelope surrounding 
the protostar. The panels (a) and
(b) show the snapshots when the stellar mass is $14.5~M_\odot$
and $16.5~M_\odot$, respectively. The elapsed time since the birth 
of the embryo protostar is also indicated on the upper-left 
corner. 
Each panel shows the spatial distributions of temperature 
({\it upper left}), density ({\it upper right}), and molecular 
fractions of hydrogen ({\it lower left}) and deuterium ({\it lower right}).    
The arrows represent the velocity field 
with different scalings between the upper and lower panels.
The open triangles in the lower panel indicate directions
only for velocities higher than 10.5 km/sec.
The dashed lines in the lower half of each panel define the
edge of the \hii region. 
}
\label{fig:snapshot}
  \end{center}
\end{figure}

After the stellar mass exceeds $10~M_\odot$, an \hii region 
begins to develop, growing in the polar directions of the disk. 
Figure~\ref{fig:snapshot}-(a) shows the gas 
distribution in the vicinity of the protostar at $M_* = 14.5~M_\odot$.
We see that the bipolar \hii region is breaking out from 
the accreting envelope. 
H$_2$ and HD molecules are completely photodissociated 
by stellar H$_2$ dissociating ($11.2~{\rm eV} \leq h \nu \leq 13.6~{\rm eV}$)
photons except in the shadow of the disk. 
Within the \hii region the ionized gas 
flows outward at a velocity of several $10$~km/s.
The high pressure in the \hii region drives a strong D-type shock,
which compresses and heats up the surrounding materials to
$\sim 10^3$~K.
The shocked gas has an outward velocity of several km/s.
The \hii region dynamically expands into the accretion
envelope and its opening angle increases with 
protostellar mass (Fig.~\ref{fig:snapshot}-b). 

\begin{figure}
  \begin{center}
\epsscale{1.1}
\plotone{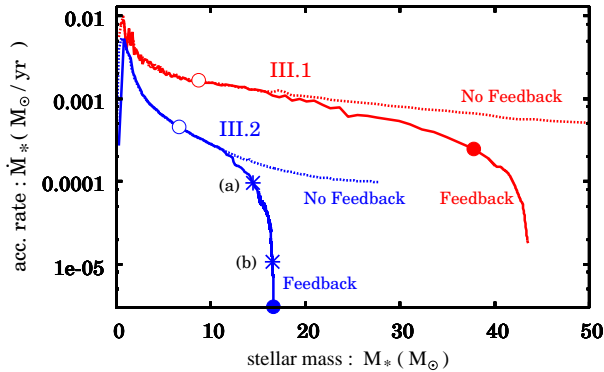}
\caption{Evolution of the mass accretion rates onto the protostar.
The blue solid curve represents the fiducial Pop III.2 case. 
The asterisks on this curve indicate the moments of the snapshots
presented in Figure~\ref{fig:snapshot}.
The open and filled circles mark the characteristic epochs as in
Figure~\ref{fig:mr_suv}.
The blue dotted curve denotes the reference case, 
whereby the stellar UV feedback was switched off. 
The red solid and dotted curves represent the fiducial Pop III.1 cases 
with and without the stellar UV feedback taken from \citet{HOYY11}.
}
\label{fig:mr_xmdot}
  \end{center}
\end{figure}

The growth of stellar mass via accretion continues even after
the formation and initial expansion of the \hii region. 
However, the accretion rate is reduced 
by radiative feedback when compared to the no-feedback case 
(Fig.~\ref{fig:mr_xmdot}). In contrast to the stellar 
UV radiation, the shock driven by the expanding \hii region 
propagates around the disk and reaches the accreting envelope 
in the disk's shadow.
The shocked gas flowing outward hinders accretion from the
envelope onto the disk.
The isolated disk exposed to the stellar UV radiation 
thins down via photoevaporation. Accretion 
from the disk onto the star finally terminates 
at $M_* \simeq 16.6~M_\odot$.
This coincides with the moment when the protostar reaches
the zero-age main-sequence stage
(Fig.~\ref{fig:mr_suv}-b).

Although the above overall evolution is qualitatively similar to 
that of Pop III.1 stars (H11), the resultant final stellar mass 
$M_* \simeq 16.6~M_\odot$ is lower than for the fiducial III.1 
case $\simeq 43~M_\odot$ (Figure \ref{fig:mr_xmdot}).
This is mostly due to the fact that the protostellar
evolution differs with the different accretion histories
between the Pop III.1 and III.2 cases (Fig.~\ref{fig:mr_suv}). 
At the lower accretion rate, in general, the KH timescale 
becomes comparable to the accretion timescale at a lower luminosity.
The protostar thus begins to lose its internal energy 
and contracts at lower stellar mass \citep[e.g.,][]{OP03,HO09}.
The stellar ionizing luminosity significantly increases during 
the KH contraction stage. Therefore, the UV feedback is effective 
at a lower stellar mass in the Pop III.2 case.  
Although the mechanism of dynamical expansion of an
\hii region and subsequent photoevaporation of the disk, which ultimately
shuts off the mass accretion, is common between
the Pop III.1 and III.2 cases, 
the required stellar ionizing luminosity for 
this mechanism to operate is much lower 
in the Pop III.2 case than for the Pop III.1 case 
(Fig.~\ref{fig:mr_suv}-b). This is because the \hii region 
expands more easily through the lower-density Pop III.2 
accretion envelope.
The accompanying protostellar outflow and photoevaporation rates
are also smaller than in the Pop III.1 case.
In the Pop III.2 case, however, the weak UV stellar feedback 
is enough to shut off the slow mass accretion 
toward the protostar.

\section{Discussion}
\label{sec:dis}

\begin{figure}
  \begin{center}
\epsscale{1.1}
\plotone{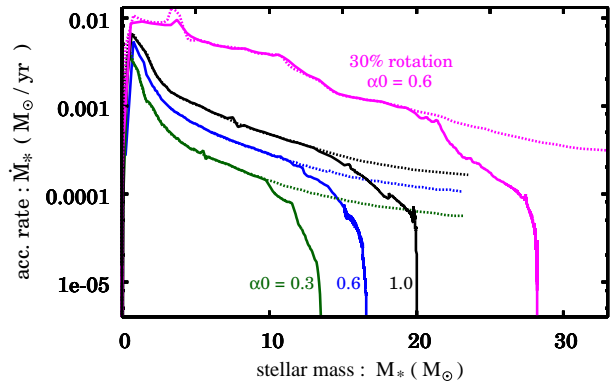}
\caption{Evolution of the mass accretion rates onto the 
protostar with different parameters.
The green, blue, and black lines represent the cases with different
$\alpha$-viscosity parameters of $\alpha_0 = 0.3$, 
0.6 (fiducial case), and 1.0. 
The magenta lines denote the test cases where the rotation
of the cloud is artificially reduced to 30\% of the
original values. The solid and dashed lines represent the evolution 
with and without the stellar UV feedback in each case. 
}
\label{fig:m_xmdot_a}
  \end{center}
\end{figure}

In this {\it Letter}, we have shown that radiative feedback 
limits the final mass of a ``typical'' Pop III.2 protostar to 
$\la 20~\msun$. 
With all the possible variations of initial conditions among
different primordial gas clouds, we naturally expect that 
the resulting final stellar masses would have some distribution.
Cosmological simulations predict that, in the Pop III.1 case, 
the accretion rate onto a star could differ by more than two orders of 
magnitude among different star-forming clouds which form through 
various halo assembly histories \citep[e.g.,][]{Osh07}.
Although such systematic studies have not been conducted for the 
Pop III.2 case, we expect a similar variance and hence some distribution 
of the final stellar masses. Here we examine potential variances of the 
final stellar masses in our simulations.

For instance, we use the $\alpha$-viscosity 
formalism \citep{SS73} for mimicking the torque exerted by 
non-axisymmetric spiral arms formed in massive self-gravitating
disks \citep[e.g.,][]{YS02,Kuiper10}. 
Recent 3D simulations show that the effective $\alpha$-parameter 
is around $0.1 - 1$ in this case \citep[e.g.,][]{Cl11,Kuiper11}; 
we adopt an $\alpha$-parameter in the mid-plane 
$\alpha_0 = 0.6$ (see H11) for the fiducial case discussed above. 
Figure~\ref{fig:m_xmdot_a} shows the accretion histories and 
resulting final stellar masses for different values 
of $\alpha_0$.
Analogously to H11, the final stellar mass increases 
with the value of $\alpha_0$: $M_\ast \simeq 13.6~M_\odot$ 
for $\alpha_0 = 0.3$, whereas
$M_\ast \simeq 20.0~M_\odot$ for $\alpha_0 = 1.0$.
Figure~\ref{fig:m_xmdot_a} also shows the cases where the initial
rotation of the natal cloud is reduced to 30\% of the fiducial value. 
For this case the stellar UV feedback operates 
only after $20~M_\odot$ of the gas has accreted onto the star and 
the resulting final stellar mass is $M_* \simeq 28~M_\odot$.

Although this mass range is systematically lower than 
that of Pop III.1 stars derived in H11,
the difference is within a factor of a few.
In reality, the existence of photodissociating 
background radiation could further erase the difference in 
the final masses of Pop III.1 and III.2 stars. 
Recall that the lower masses of Pop III.2 stars originates 
from the enhanced abundances of H$_2$ and HD, 
and the resultant more efficient cooling during the
pre-stellar collapse of the natal cloud.
However, the presence of even a weak photodissociating
background ($\sim 10^{-22} {\rm erg~cm^{-2}~sr^{-1}~sec^{-1}~Hz^{-1}}$) 
inhibits the formation of 
H$_2$ and HD molecules, thereby making the thermal 
evolution almost the same in the 
Pop III.1 case \citep[e.g.,][]{YOH07,WGH11}.
If this is the case, the mass scale of Pop III.1 and III.2 stars 
would be identical.

Although not considered in this study, gravitational
fragmentation of the disk would also contribute to shape the
mass distribution of primordial stars.
For the Pop III.1 case, recent 3D numerical simulations
show that the circumstellar disk easily becomes gravitationally
unstable to fragmentation \citep[e.g.,][]{Mcd08,Stacy10,Cl11,Gr12}. 
However, most of these studies consider the early evolution only,
far before the mass accretion ceases and final stellar masses 
are determined. 
One can naively expect that disk fragmentation reduces the final 
stellar mass as the accreting material is shared by multiple 
protostars born from the fragments
\citep[so-called ''fragmentation-induced starvation'',
e.g.,][]{Peters10}.

The nature of mass accretion with disk fragmentation 
might be even more complex during later evolutionary phases. For instance,
gravitational interactions between the multiple protostars could
redistribute angular momentum.  If angular momentum is carried away
by a small number of (proto-)stars thrown out of the system, accretion
of the remaining gas onto the remaining star(s) could be enhanced.  
Moreover, merging of protostars in the disk is also possible.  The net
result is that mass accretion onto the star would be strongly
time-dependent or even have the consequence of stochastic burst-like
accretion events which potentially change the evolution of the
protostars \citep[but see, e.g.,][]{Smith12}. 
These effects should be examined in future 
long-term 3D simulations.

If some of the low-mass stars born via the disk fragmentation 
survive without significant mass growth, they would be low-mass 
satellite-like stars orbiting around the massive stars whose 
final masses are limited by stellar radiative feedback.
Future work would explain how this effect differs 
between the Pop III.1 and III.2 cases \citep[e.g.,][]{Mcd09}.
As shown in \citet{Cl11}, disk fragmentation is 
controlled by a balance between the mass transfer rates onto and 
through the disk.
Systematic difference in accretion rates onto the star-disk systems 
among the Pop III.1 and III.2 cases  
would result in basic differences in the effects of disk fragmentation.

Our results provide a new picture that the typical mass of the
primordial stars are tens of solar masses regardless
of their generations. This explains why the signatures of 
pair-instability supernova, which is the final fate of 
$140~M_\odot \lesssim M_* \lesssim 260~M_\odot$ stars, have
not been seen in the abundance patterns of 
Galactic metal-poor stars \citep[e.g.,][]{TVS04,FJB09,Caffau11}.

{\acknowledgements 
The authors thank Neal Turner and Rolf Kuiper
for fruitful discussions and comments.
T.H. appreciates the support by Fellowship of the Japan
Society for the Promotion of Science for Research Abroad.
K.O. is supported by the Grants-in-Aid by the Ministry of 
Education, Science and Culture of Japan (2168407 and 21244021).
Portions of this work were conducted at the Jet Propulsion Laboratory,
California Institute of Technology, operating under a contract with 
the National Aeronautics and Space Administration (NASA).}

\end{document}